# Model Free Deep Deterministic Policy Gradient Controller for Setpoint Tracking of Non-minimum Phase Systems

Fatemeh Tavakkoli, Pouria Sarhadi, Benoit Clement, Wasif Naeem

*Abstract—* Deep Reinforcement Learning (DRL) techniques have received significant attention in control and decision-making algorithms. Most applications involve complex decision-making systems, justified by the algorithms' computational power and cost. While model-based versions are emerging, model-free DRL approaches are intriguing for their independence from models, yet they remain relatively less explored in terms of performance, particularly in applied control. This study conducts a thorough performance analysis comparing the data-driven DRL paradigm with a classical state feedback controller, both designed based on the same cost (reward) function of the linear quadratic regulator (LQR) problem. Twelve additional performance criteria are introduced to assess the controllers' performance, independent of the LQR problem for which they are designed. Two Deep Deterministic Policy Gradient (DDPG)-based controllers are developed, leveraging DDPG's widespread reputation. These controllers are aimed at addressing a challenging setpoint tracking problem in a Non-Minimum Phase (NMP) system. The performance and robustness of the controllers are assessed in the presence of operational challenges, including disturbance, noise, initial conditions, and model uncertainties. The findings suggest that the DDPG controller demonstrates promising behavior under rigorous test conditions. Nevertheless, further improvements are necessary for the DDPG controller to outperform classical methods in all criteria. While DRL algorithms may excel in complex environments owing to the flexibility in the reward function definition, this paper offers practical insights and a comparison framework specifically designed to evaluate these algorithms within the context of control engineering.

I. INTRODUCTION

Model-free Deep Reinforcement Learning (DRL) algorithms have emerged in various domains, including gaming [1,2] and robotic control [3]. The promise of automating a wide range of decision-making, with a particularly strong connection to control and adaptive systems, is shared by the combination of Reinforcement Learning (RL) with high-capacity function approximators like neural networks [4]. Unlike supervised and unsupervised learning paradigms, RL entails learning a policy (by agent or controller) through the optimization of a reward function (control objective) via interaction with the environment (system), utilizing states (measurements) and actions (control signals) [5].

The challenge of enhancing the generalizability of RL has been significantly investigated over the past decade, marking a turning point with the introduction of Deep Q-Network (DQN) and continuous action approaches such as Deep Deterministic Policy Gradient (DDPG). DQN employs a deep neural network as a function approximator of the value function [6]. DQN addresses the inherent instability problem associated with function approximation in reinforcement learning by utilizing two techniques: experience replay [7] and target networks [8]. Deterministic policy gradient algorithms represent another advancement in RL algorithm, extending the standard policy gradient theorems from stochastic policies to deterministic policies [9]. Notably, three well-known algorithms in this category are DDPG [10], twin Delayed Deep Deterministic Policy Gradient (TD3) [11], and Proximal Policy Optimization (PPO) [12]. Initially developed using Python code, the remarkable level of acceptance and demand for these algorithms has even led to their integration into MATLAB toolboxes. Those approaches offer attractive model-free solutions that have received notable attention in control and decision-making (mission planning) systems [13]-15]. However, their performance, robustness, and limitation boundaries have not been thoroughly identified and analyzed to date. This is despite the fact that even well-established algorithms like Smith Predictors, with longstanding history, are now understood as not superior to a PI controller [16]. Therefore, a meticulous analysis of these approaches is necessary for their utilization and future developments.

A few studies have explored the application of these methods in process control applications, comparing their performance against control system requirements. For example, references [17-18] have examined the application of DDPG, TD3, PPO, TRPO, and SAC in specific process control systems, which is commendable. Superior performance for DDPG-based algorithms has been reported [17]. However, there remains a gap in the analysis of these

Fatemeh Tavakkoli was with the Faculty of Electrical and Computer Engineering Department, Babol Noshirvani University of Technology, Babol, Iran (e-mail: fatemetavakkolii71@gmail.com)
Pouria Sarhadi is with the School of Physics, Engineering and Computer Science, University of Hertfordshire, Hatfield, AL10 9AB, Hertfordshire, UK (phone: +44(0)1707 285961; e-mail: p.sarhadi@herts.ac.uk)
Benoit Clement is with ENSTA Bretagne, CROSSING ILR CNRS 2010 and Flinders University, Australia (e-mail: benoit.clement@ensta-bretagne.fr)
Wasif Naeem is with the School of Electronics, Electrical Engineering and Computer Science, Queen's University Belfast, UK, (e-mail: w.naeem@qub.ac.uk).

algorithms' performance in practical control challenges, including non-minimum phase (NMP) behavior, disturbances, noise, initial conditions, and model uncertainties such as delay. Additionally, analyzing the quality of the control signals is another crucial aspect to gain insights into their effectiveness in control applications.

This study is dedicated to conducting a comprehensive assessment of DDPG's performance in demanding scenarios, particularly in comparison to a classical optimal controller, the Linear Quadratic Integral (LQI) controller, with a focus on the setpoint tracking problem in NMP systems. Given the widespread use of DDPG, we have selected it for the performance evaluation and gathering insights that can be applied to other applications. Two DDPG controller variants have been developed: one estimates the state feedback controller gains, while the other functions as a black-box data-driven controller without any model structure. It is important to note that this study does not advocate for the replacement of well-established model-based controllers with computationally intensive techniques like DDPG. Instead, it offers quantitative insights into model-free DRL for potential future enhancements. As a contribution, the paper introduces a comprehensive framework with detailed criteria for evaluating practicality and facilitating comparative analyses of DRL techniques.

## II. THE ANALYSIS FRAMEWORK

Twelve essential design criteria have been chosen to assess the controllers' performance. They encompass critical aspects, including transient and steady state performance, cumulative error in step response tracking, robustness, and the quality of the control signal. They are explained as follows:

**C1.** *Rise time* ($t_r$) is defined as the duration it takes for the response to increase from 0 to 100 percent of its ultimate value.

**C2.** *Maximum Overshoot* ($M_p$) is the highest peak value of the response curve, as measured from the system's desired response. For a step input, the percentage overshoot ($M_p$) is considered.

**C3.** *Maximum Undershoot* ($M_u$), similar to overshoot but in the opposite direction, quantifies the maximum dip in the signal as a percentage of the desired response. Undershoot is an inherent phenomenon in these systems, unavoidable, but its peak value should be fine-tuned for high-performance tracking.

**C4.** *Settling Time* ($t_s$) denotes the duration required for a system's response to reach and stay within a predefined range threshold ($e_{ss}$) around its final steady-state value.

**C5.** *Steady State error* ($e_{ss} = r - y(t_f)$) is the difference between the desired or reference output (r) and the actual output (y) of a system in its steady-state condition.

**C6.** Integral of the Square of the Error (ISE) is another traditional performance metric that assesses the cumulative square of the error between the system's output and the desired output over time (from $t_0 = 0$ to $t_f$). It emphasizes both the magnitude and the duration of the error:

$$ISE = \int_0^{t_f} e^2(t)dt \qquad (1)$$

**C7.** Integral of the Time multiplied by Absolute Error ($ITAE$) is another cumulative performance criterion that places larger weights for steady-state error by an increasing time weight:

$$ITAE = \int_0^{t_f} t|e(t)|\, dt \qquad (2)$$

Both *ISE* and *ITAE* are standard indices in control systems' tracking performance [19].

**C8.** Integral of the Absolute of the Control Effort ($IACE$) quantifies the cumulative absolute of the control effort or control signal ($u_c$) applied over time, representing consumed energy. It can be calculated using the following formula:

$$IACE = \int_0^{t_f} |u_c(t)|dt \qquad (3)$$

**C9.** Integral of the Absolute Control Effort Rate ($IACER$) measures the integral of the absolute rate of change of the control effort or control signal applied to the system:

$$IACER = \int_0^{t_f} |du_c(t)|dt \qquad (4)$$

Maintaining lower bounds on control signal variations is pivotal for preventing rate saturation within actuation systems and its adverse effects on closed-loop stability [20].

**C10.** Control Signal Maximum Value ($u_{cmax}$) represents the maximum value or amplitude of the control signal applied to the system as follows:

$$u_{cmax} = \max_t |u_c(t)| \qquad (5)$$

To prevent amplitude saturation in actuators, keeping $u_{cmax}$ below the input limitations of the system is advisable.

*IACE*, *IACER*, and $u_{cmax}$ indices play a crucial role in designing a practical controller. Most of the time, a controller that can satisfy tracking performance (C1-C7) while minimizing control quality indices (C8-C10) is preferred but it can vary depending on the application case.

**C11.** Gain Margin ($GM$) is a frequency domain stability measure for a linear closed-loop system ($G_{CL}$). It quantifies the range of adjustability in the open-loop system ($G_{OL}$) gain before instability occurs. GM is calculated as [19]:

$$GM = 20\, log\left(\frac{1}{|G(j\omega_{pc})|}\right) \qquad (6)$$

where is $|G(j\omega_{pc})|$ the magnitude of the system's transfer function at the phase crossover frequency $\omega_{pc}$ where phase shift is -180 degrees. A larger $GM$ indicates greater stability.

**C12.** Delay Margin ($DM$), as a practical stability index, represents the time delay margin of a system and indicates the maximum value of time delay that can be introduced to $G_{OL}$ without causing instability in the closed-loop system $G_{CL}$ [21].

It is measured in seconds and can theoretically be calculated using the following formula:
$$DM = \frac{PM}{|G(j\omega_{gc})|} \quad (7)$$
where $PM$ is the system phase margin and $\omega_{gc}$ symbolizes the gain crossover frequency [21]. We appreciate that both $GM$ and $DM$ are frequency domain indicators, which may not be applied to ML-based or nonlinear loops. However, one can manually increase the amount of gain or delay uncertainty to determine these margins, providing valuable insights into the system's stability.

Together, the introduced criteria (C1-C12) can serve as valuable tools for assessing and comparing controller performance. Hence, we suggest using them as standard comparison metrics in control system design.

## III. LINEAR QUADRATIC CONTROLLER WITH AUGMENTED ERROR INTEGRAL STATE FEEDBACK (LQI)

This Section provides a brief description of the LQI controller as the classic baseline for comparison [22]. The reasons to select this controller are: i) it is a classical controller with wide use of applications inheriting PID-like performance in lower order systems, ii) incorporating a cost function comprising tracking error and the control signal that can be directly used as a reward function for a fair comparison with DPPG. Consider a plant with the following state-space representation:
$$\dot{x}_p = A_p x_p + B_p u$$
$$y = C_p x_p \quad (8)$$
Where $x_p$, $u$, and $y$ denote state vector, control signal, and output respectively. The matrices $(A_p, B_p, C_p)$ describe the dynamics of the plant, which does not necessarily include tracking. By introducing a new integrator state where $(e_{yI})$ is the integral of the error, the $\dot{e}_I = e$ a tracking error, an augmented system with states $x = [x_p \quad e_I]^T$ can be achieved as follows [19]:
$$\underbrace{\begin{bmatrix} \dot{x}_p \\ \dot{e}_I \end{bmatrix}}_{\dot{x}} = \underbrace{\begin{bmatrix} A_p & 0 \\ -C_p & 0 \end{bmatrix}}_{A} \underbrace{\begin{bmatrix} x_p \\ e_I \end{bmatrix}}_{x} + \underbrace{\begin{bmatrix} B_p \\ 0 \end{bmatrix}}_{B} u + \underbrace{\begin{bmatrix} 0 \\ I \end{bmatrix}}_{B_r} r \quad (9)$$
Where $A$, $B$, and $C$ are the augmented system's matrices. Utilizing a quadratic cost function as follows:
$$J = \frac{1}{2}\int_0^\infty (x^T Q_{LQR} x + u_c^T R_{LQR} u_c) dt \quad (10)$$
It is shown by solving the following Riccatti equation [22]:
$$A^T P + PA + Q_{LQR} - PBR_{LQR}^{-1}B^T P = 0 \quad (11)$$
The following control signal can be achieved:
$$u_c(t) = -K_{LQR} x_p(t) + k_I e_I(t) \quad (12)$$
Fig. 1 illustrates the block diagram of LQI controller. This controller inherits PID attributes with optimal gains calculated using LQR which makes it popular in realistic applications.

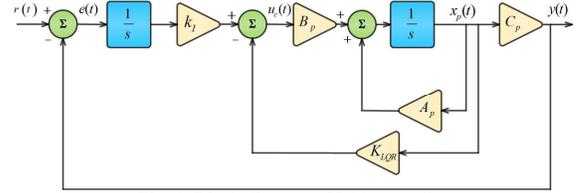

Fig. 1: Block diagram of LQI controller

## IV. DETERMINISTIC POLICY GRADIENT AS A MODEL FREE SETPOINT TRACKER

Another way to tune a controller is based on RL. It revolves around the concept of an agent engaging with its environment, aiming to learn behavior that maximizes reward [10]. At each discrete time step, denoted as $t$, the agent operates in a specific state ($s \in S$) and selects actions ($a \in A$) based on a policy ($\pi: S \to A$). Consequently, the agent receives a reward $r$ and the environment transitions to a new state $s'$.

The primary objective in RL is to identify the optimal policy $\pi_\phi$, with parameters $\phi$, that maximizes the expected return $J(\phi) = E_{s_i \sim p_\pi, a_i \sim \pi}[R_0]$. For continuous control, parametrized policies $\pi_\phi$ can be updated by taking the gradient of the expected return $\nabla_\phi J(\phi)$. In actor-critic methods, the policy, referred to as the actor, can be updated using the deterministic policy gradient algorithm:
$$\nabla_\phi J(\phi) = E_{s \sim p_\pi}[\nabla_a Q^\pi(s,a)|_{a=\pi(s)} \nabla_\phi \pi_\phi(s)] \quad (13)$$
$Q^\pi(s,a) = E_{s_i \sim p_\pi, a_i \sim \pi}[R_t|s,a]$, the expected return when performing action $a$ in state $s$ and following $\pi$ afterward, is known as the critic or the value function. In this paper, two DDPG agents are developed to approximate the discounted long-term reward through a Q-value-function critic. In the first algorithm ($DDPG_1$), the critic network endeavors to estimate state feedback gains referred to as $K_{DDPG}$, the same structure employed in the LQI controller. The second method ($DDPG_2$) deploys an end-to-end black-box controller that maximize the reward function. To facilitate a comprehensive analysis and comparison with the LQI controller, both methods utilize identical observations represented by $s = x = [x_1 \quad x_2 \quad x_3]^T = [x_p \quad e_I]^T$. The reward equation is defined as the negative cost function in LQR as below:
$$r(t) = -(x(t)^T Q_{LQR} x(t) + u_c R_{LQR} u_c) \quad (14)$$
In $DDPG_1$, the critic implemented as a neural network, accepts observations, and actions as inputs and returns scalar values. It utilizes a network with both quadratic and fully connected layers to approximate the quadratic value function of the optimal policy. The policy implemented by the actor, which has been selected by $a = u_c = K_1 x_1 + K_2 x_2 + K_3 x_3 = K_{DDPG} x$. The matrix form of the Q-value function is defined as follows Where $w_i$ are the weights of the fully connected layer:

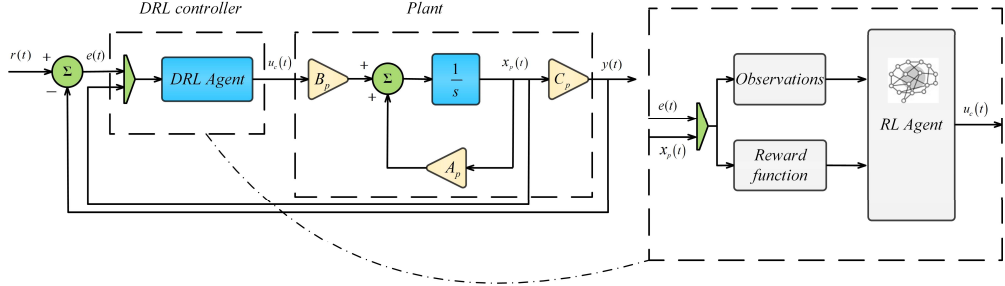

Fig. 2: Block diagram of the setpoint tracking DDPG controller

$$Q(x, u_c) = [x_1 \quad x_2 \quad x_3 \quad u_c] \begin{bmatrix} w_1 & \frac{w_2}{2} & \frac{w_3}{2} & \frac{w_4}{2} \\ \frac{w_2}{2} & w_5 & \frac{w_6}{2} & \frac{w_7}{2} \\ \frac{w_3}{2} & \frac{w_6}{2} & w_8 & \frac{w_9}{2} \\ \frac{w_4}{2} & \frac{w_7}{2} & \frac{w_9}{2} & w_{10} \end{bmatrix} \begin{bmatrix} x_1 \\ x_2 \\ x_3 \\ u_c \end{bmatrix} \quad (15)$$

For a fixed policy $u_c = K_{DDPG} x(t)$, the cumulative long-term reward is:

$$R(x) = Q(x, K_{DDPG} x) = x^T [I \quad K_{DDPG}^T] w \begin{bmatrix} I \\ K_{DDPG} \end{bmatrix} x \quad (16)$$
$$= x^T P x$$

Since the rewards are always negative, to properly approximate the cumulative reward both $P$ and $W$ should be negative definite.

The $DDPG_2$ algorithm combines the strengths of policy-based and value-based methods by exploiting two neural networks: the actor-network determines the optimal actions given the current state, while the critic-network estimates the state-action value function (Q-function), which represents the expected cumulative reward. Fig. 2 shows the block diagram of the developed DDPG set-point tracking controller.

## V. SIMULATION RESULTS

The LQI and DDPG controllers are simulated on a NMP system tracking problem, and their behavior is analyzed using the proposed framework in Section II. The considered plant's transfer function is as follows:

$$G_P = \frac{0.5s - 1}{s^2 + 3s + 2} \quad (17)$$

The system is stable but exhibits NMP behavior due to the presence of a right-hand side zero. Challenges associated with controlling such systems are well-recognized within the control community [23]. These challenges result in an undershoot in the step response and can pose limitations for achievable performance in control. Thus, we explore the performance of the model-free DRL approach in addressing such systems. Its controllability state-space representation can be expressed as follows:

$$A_p = \begin{bmatrix} 0 & 1 \\ -2 & -3 \end{bmatrix}, \quad B_p = \begin{bmatrix} 0 \\ 1 \end{bmatrix}, \quad C_p = [-1 \quad 0.5] \quad (18)$$

The parameters of the reward function are selected as:

$$Q_{LQR} = diag(0,0,10), \quad R_{LQR} = 1 \quad (19)$$

where $diag$ is a diagonal matrix with weight 10 for tracking error and $R_{LQR} = 1$ as the control weight to achieve a tradeoff between tracking and energy consumption. The DDPG algorithms employ two networks: a critic network and an actor network. The critic network takes observations and actions as inputs and generates output values through layers. The actor network employs a parametrized continuous deterministic policy to produce corresponding actions based on observations. In the critic network of $DDPG_1$, the bias learn rate factor is set to 4 with a bias value of zero, while both parameters are set to zero in the actor network. In $DDPG_2$, the critic estimates the state-action value function using two fully connected layers followed by a rectified linear unit (ReLU) layer for observations, and one fully connected layer for actions. After combining results of these layers there are two fully connected layers and two ReLU layers, each with a hidden layer size of 100 for all fully connected layers. The action network comprises four fully connected layers and three ReLU layers with one $tanh$ layer and scaling layer where the scale is set to 4 to determine the optimal action. The remaining agent hyper-parameters are listed in Table I. The learning curves for 850 episodes, as depicted in Fig. 3, demonstrate the agents' efforts to maximize reward.

TABLE I Hyper-parameters for DDPG controller

| Parameter | Value |
|---|---|
| Discount factor | 1 |
| Initial learn rate of critic | 1e-3 |
| Initial learn rate of actor | 1e-4 |
| Exploration noise variance | 0.1 |
| Noise variance decay rate | 1e-6 |
| Integral gain | 0.1 |
| Simulation time step | 0.1 |
| Batch size | 256 |
| Maximum steps in an episode | 200 |
| Maximum episode | 850 |

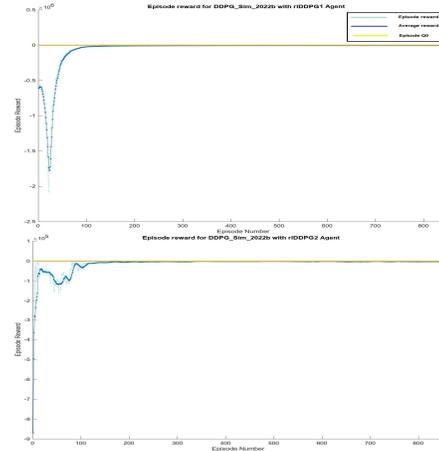

Fig. 3: Learning curves for $DDPG_1$ (top) and $DDPG_2$ (bottom)

Figure 4 illustrates the performance comparison between LQR-based and DRL-based controllers in a step setpoint tracking scenario, alongside the corresponding performance criteria listed in Table II. While all controllers achieve the tracking mission, they exhibit distinct strengths and weaknesses when evaluated against the performance criteria. Both DDPG controllers indicate a faster rise-time, however, $DDPG_1$ has higher values for $M_p$, $M_u$, and $t_s$, and $DDPG_2$ has higher $M_u$ but its $M_p, t_s$ are less than the LQI controller. $DDPG_2$ shows lower cumulative error measures than LQI, persisting throughout operation, while LQI initially experiences higher errors but improves over time. DDPG controllers demand higher maximum control effort, while $DDPG_2$ operates within lower signal boundaries. In terms of robustness, DDPG controllers demonstrate smaller gain margins and delay margins compared to the LQI controller. One can say $DDPG_2$ exhibits the best overall performance.

In the second test, we evaluate the controllers' performance under disturbance and noise conditions. An output disturbance with an amplitude of 0.2 and white noise with a standard deviation of 0.1 are introduced at the 15$^{th}$ and 20$^{th}$ seconds, respectively. The response is shown in Fig. 5, and performance comparison for the affected criteria are presented in Table III. The results suggest that the DRL controller performs adequately under perturbed conditions and slightly outperforms LQI controller. These results are intriguing as the DRL controller was not specifically trained for the perturbed condition.

To analyze performance scalability, Figure 6 displays results for two different initial conditions for $DDPG_2$ and LQI ($x_0 = [1; -2], x_0 = [-1; 2]$). $DDPG_2$ demonstrates smaller undershoot and higher overshoot compared to LQI. Both controllers have same rise time while the $DDPG_2$ exhibits a slightly shorter settling time. While scalability for a linear controller like LQI is predictable, these tests highlight DRL's performance in challenging control problems. It is important to note that none of the controllers can be definitively deemed superior; however, the performance of $DDPG_2$ is partially better. Nonetheless, considering the tuning efforts and computational burden required for DDPG, LQI may be preferred for systems with an available model.

TABLE III Performance criteria of the controllers in perturbed condition

| No. | Criteria | LQI | $DDPG_1$ | $DDPG_2$ |
|---|---|---|---|---|
| 1 | ISE | 21.2 | 22.3 | 14.1 |
| 2 | ITAE | 268.3 | 297.6 | 170.7 |
| 3 | IACE | 446.6 | 477.2 | 481.4 |
| 4 | IACER | 72.7 | 237.6 | 62.8 |

TABLE II Performance criteria of the controllers in nominal tracking

| No. | Criteria | LQI | $DDPG_1$ | $DDPG_2$ |
|---|---|---|---|---|
| 1 | $t_r$ | 3.0 | 2.5 | 1.7 |
| 2 | $M_p$ | 21.5% | 45.1% | 15.3% |
| 3 | $M_u$ | 7.7% | 7.7% | 22.5% |
| 4 | $t_s$ | 6.7 | 9.4 | 3.7 |
| 5 | $e_{ss}$ | 0 | 0 | 0 |
| 6 | ISE | 19.4 | 20.9 | 12.8 |
| 7 | ITAE | 52.2 | 121.9 | 12.6 |
| 8 | IACE | 406.8 | 419.3 | 420.3 |
| 9 | IACER | 40.4 | 82.7 | 26.9 |
| 10 | $u_{c\,max}$ | 2.8 | 3.4 | 4.00 |
| 11 | GM | 27.8 | 2.3 | 18.1 |
| 12 | DM | 0.85 | 0.08 | 0.55 |

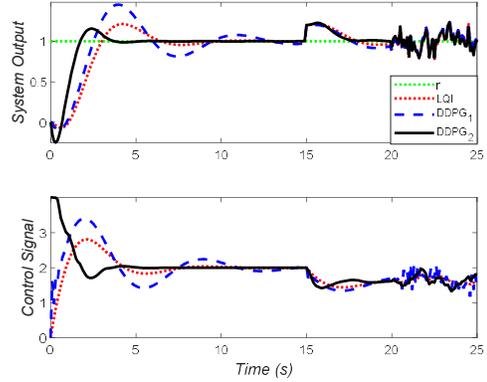
Fig. 5: Setpoint tracking in the presence disturbance and noise

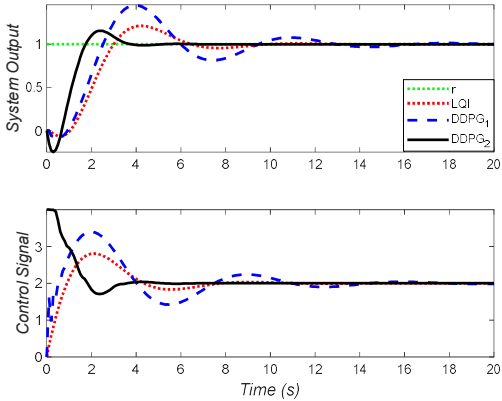
Fig. 4: Step response for DDPG and LQI controllers

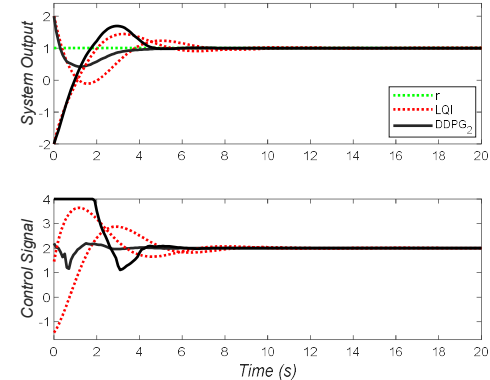
Fig. 6: Step response in the presence of initial conditions

## VI. Conclusion

The paper offers a transparent comparison between a state-of-the-art DRL-based algorithm (DDPG) and a well-established classical controller (LQI). This comparison aims to improve the understanding of how emerging RL techniques perform in control applications. While DDPG and its model-free counterparts are recently applied to numerous decision-making and control problems, elaborated and explicit comparison are lacking. For this purpose, a NMP system setpoint tracking problem is considered in various testing scenarios. From a control standpoint, twelve tailored criteria are proposed to quantify performance across various critical aspects. These aspects include transient and steady-state performance, cumulative error in tracking, robustness, and the quality of the control signal. The proposed framework can be adopted in similar studies. Findings are summarized below:

- The DDPG-based controller can achieve setpoint tracking performance for NMP systems in the presence of disturbances, noise, and uncertainties in gain and delay, performing comparably or better than the classical LQI controller.
- While promising, model-free DDPG indicates lower robustness ($GM$ and $DM$) under similar conditions.
- Despite not requiring a model and featuring reasonable training time, the DRL controller is highly sensitive to tuning parameters. Slight variations in hyperparameters can lead to successful convergence or poor performance with the same reward function. The agent's performance is significantly affected by its training parameters, highlighting a key area for future investigation.
- In general, despite appealing features, the DRL controller cannot showcase a considerably better performance. This follows a similar trend as reported in other applications, such as signal processing [24].
- One notion entails integrating model-based controllers and DRL techniques [25], a paradigm that necessitates employing or estimating models even with reduced precision, which is beyond the scope of this research.
- Our findings are constrained by the existing benchmark; DRL solutions may show superior performance in other complex applications. However, this requires further studies. Nonetheless, similar behavior has been observed in other analogous control problems.
- Further evaluations are recommended to assess the performance and robustness of existing DRL controllers. Employing a fair and transparent methodology, similar to the one used in this paper, is advisable. Ongoing efforts are crucial to enhancing the performance of current algorithms.
- Finally, our simulations rely on the MATLAB RL toolbox codes, thus its validity is an assumption.

Overall, promising performance is observed with DDPG, although it is not deemed extraordinary at present and has potential for improvement in the future. This study does not advocate for the replacement of well-established classical controllers like state feedback LQI with computationally intensive techniques such as DDPG. Rather, its objective is to provide a quantitative analysis of model-free data-driven DRL controllers for potential future enhancements.